\documentclass[aps,prd,twocolumn,superscriptaddress,preprintnumbers,floatfix,showpacs,nofootinbib]{revtex4-1}

\usepackage{graphicx}
\usepackage[caption=false]{subfig}
\usepackage{amsmath}
\usepackage{amssymb}
\usepackage{tikz}
\usepackage{siunitx}
\usepackage{color}
\usepackage[colorlinks=True]{hyperref}
\usepackage{standalone}
\usepackage{microtype}

\usepackage{draftwatermark}
\SetWatermarkScale{6}

\usetikzlibrary{calc}

\usetikzlibrary{patterns}

\usepackage{etoolbox} 

\makeatletter
\def\watermarkoff{%
 \@sc@wm@stampfalse
}
\makeatother

\makeatletter
\def\watermarkon{%
 \@sc@wm@stamptrue
}
\makeatother

\newcommand{\beq}{\begin{equation}}
\newcommand{\eeq}{\end{equation}}

\newcommand*{\eq}[1]{Eq.\ \eqref{eq:#1}}
\newcommand*{\fig}[1]{Fig.\ \ref{fig:#1}}

\newcommand*{\hyp}[1]{{{\cal H}_{\rm #1}}}

\newcommand{\fvec}[1]{\vec{#1}}

\newcommand*{\blue}[1]{#1}

\newcommand{\dcc}{LIGO-P1700276}

\graphicspath{{./fig/}}

\begin{document}

\preprint{\dcc}

\title{Probing gravitational wave polarizations with \\signals from compact
binary coalescences}


\author{Maximiliano Isi}
\email[]{misi@ligo.caltech.edu}
%\thanks{}
\affiliation{LIGO Laboratory, California Institute of Technology,
Pasadena, California 91125, USA}

\author{Alan J.\ Weinstein}
%\email[]{ajw@ligo.caltech.edu}
%\thanks{}
\affiliation{LIGO Laboratory, California Institute of Technology,
Pasadena, California 91125, USA}


\date{\today}

\begin{abstract}
In this technical note, we study the possibility of using networks of
ground-based detectors to directly measure gravitational-wave polarizations
using signals from compact binary coalescences. We present a simple data
analysis method to partially achieve this, assuming presence of a strong signal
well-captured by a GR template.
\end{abstract}

% \pacs{04.80.Cc, 04.30.Nk, 04.50.Kd, 04.80.Nn I.}
%\keywords{}

\maketitle

\watermarkoff

% body of paper here - Use proper Sec.\ commands

%\tableofcontents

\twocolumngrid
\section{Introduction}

The detection of gravitational waves (GWs) by the Advanced Laser Interferometer
Gravitational-Wave Observatory (aLIGO) has enabled some of the first
experimental studies of gravity in the highly dynamical and strong-field
regimes \cite{gw150914, gw151226, o1bbh, gw170104, gw150914_tgr}. These first
few detections have already been used to place some of the most stringent
constraints on deviations from the general theory of relativity (GR) in this
domain, which is inaccessible to laboratory, Solar System or cosmological tests
of gravity.

However, it has not been possible to use LIGO signals to learn about the
polarization content of GWs \cite{gw150914_tgr}, a measurement highly relevant
when comparing GR to many of its alternatives \cite{tegp, Will2006}. In fact,
all existing observations are so far consistent with the extreme case of purely
non-GR polarizations. The reason for this is that the two LIGO instruments are
nearly coaligned, meaning that they are sensitive to approximately the
same linear combination of polarizations. This makes it nearly impossible to
unequivocally characterize the polarization content of transient GW signals
like the compact-binary coalescences (CBCs) observed so far, at least not
without making assumptions about the way the signals were sourced
\cite{Will2006, Chatziioannou2012}.

Existing observations that are usually taken to constrain the amount of allowed
non-GR polarizations can do so only in an indirect manner. For example,
measurements of the orbital decay of binary systems are sensitive to the total
radiated GW power, but do not probe the geometric effect (namely, the
directions in which space is stretched and squeezed) of the waves directly (see
e.g.\ \cite{Weisberg2010, Freire2012}, or \cite{Stairs2003, Wex2014} for
reviews). In the context of specific alternative theories (e.g.\ scalar-tensor)
such observations can indeed constrain the power contained in extra
polarizations. However, such measurements provide no direct, model-independent
information on the actual polarization content of the gravitational radiation.
Thus, there may be multiple theories, with different polarization content, that
still predict the correct observed GW emitted power.

To see that the above is the case, consider a scenario in which GWs are emitted
precisely as in GR, but where the polarizations change during propagation: the
phase evolution would be similar to GR, but the geometric effect of the wave
would be completely different \cite{Berezhiani2007, Hassan2013, Max2017,
Brax2017}. (This polarization mutation could take place if the linear
polarization basis does not diagonalize the kinetic matrix of the theory, as is
the case for neutrino oscillations \cite{Pontecorvo1957, Pontecorvo1967}, or
for the circular GW polarization states in dynamical Chern-Simons gravity
\cite{Alexander2009}.) Because the same limitations of pulsar binary analyses
apply to studies of the details in the phasing of signals previously detected
with LIGO, and other traditional tests of GR (like Solar System tests) have no
bearing on GWs, there currently exist no direct measurements of GW
polarizations.

Prospects for the direct measurement of GW polarizations are improved by the
addition of Advanced Virgo to the detector network. In principle, at least five
noncoaligned differential-arm detectors would be needed to break {\em all} the
degeneracies among the five nondegenerate polarizations allowed by generic
metric theories of gravity \cite{Eardley1973a, Eardley1973b}, if transient
signals are used \cite{Isi2017, Callister2017}. However, as we will show, the
current Advanced-LIGO--Advanced-Virgo network can already be used to
distinguish between {\em some} of the possible combinations of polarizations
without the need to use specific knowledge about the phase evolution of the
source.

In this note, we present a simple Bayesian method to extract information about
GW polarizations directly from strong CBC signals by using the relative
amplitudes and timing at the different detectors.

\section{Background} \label{sec:background}

\subsection{Polarizations} \label{sec:polarizations}

\begin{figure}
\includegraphics[width=0.33\columnwidth]{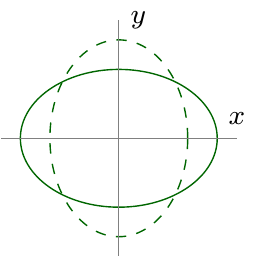}\hfill
\includegraphics[width=0.33\columnwidth]{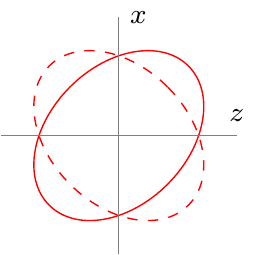}\hfill
\includegraphics[width=0.33\columnwidth]{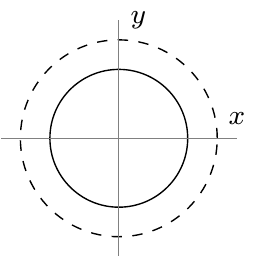}\\
\includegraphics[width=0.33\columnwidth]{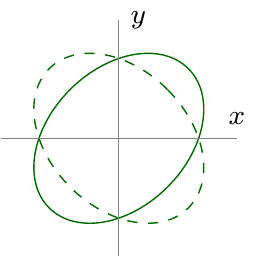}\hfill
\includegraphics[width=0.33\columnwidth]{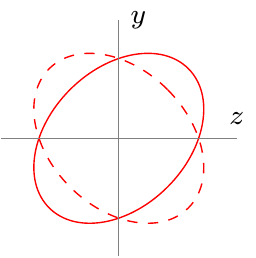}\hfill
\includegraphics[width=0.33\columnwidth]{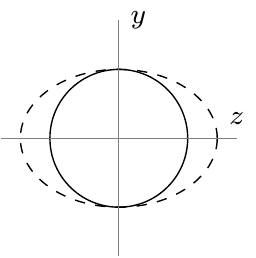}
% \includestandalone[width=0.33\columnwidth]{tikz/polcircles_p}\hfill
% \includestandalone[width=0.33\columnwidth]{tikz/polcircles_x}\hfill
% \includestandalone[width=0.33\columnwidth]{tikz/polcircles_b}\\
% \includestandalone[width=0.33\columnwidth]{tikz/polcircles_c}\hfill
% \includestandalone[width=0.33\columnwidth]{tikz/polcircles_y}\hfill
% \includestandalone[width=0.33\columnwidth]{tikz/polcircles_l}
\caption{{\em Effect of different GW polarizations on a ring of free-falling
test particles}. Plus (+) and cross ($\times$) tensor modes (green); vector-x
(x) and vector-y (y) modes (red); breathing (b) and longitudinal (l) scalar
modes (black). In all of these diagrams the wave propagates in the
\emph{z} direction. This decomposition into polarizations was first proposed
for generic metric theories in \cite{Eardley1973b}.}
\label{fig:circles}
\end{figure}

In all theories that respect Einstein's equivalence principle, including GR,
gravitational interactions may be fully described via the universal coupling of
matter to a metric tensor \cite{Thorne1973, tegp}. Because of this, it may be
shown that, in any such {\em metric theory}, a (nearly-)null plane GW may be
encoded in at most six independent components of the Riemann tensor at any
given point in spacetime \cite{Eardley1973a, Eardley1973b, tegp}. These degrees
of freedom give rise to six geometrically distinct polarizations, corresponding
to the six linearly independent components of an arbitrary metric perturbation.

At any given spacetime point $\fvec{x}$, the metric perturbation may thus be
written as
\beq \label{eq:hab}
h_{ab}(\fvec{x}) = h_{A}(\fvec{x})\, e^A_{~ab}\, ,
\eeq
for six independent amplitudes, $h_{A}(\fvec{x})$, and six polarization tensors
$e^A_{~ab}$ (implicit sum over polarizations $A$). For instance, letting
${\bf w}_z={\bf w}_x \times {\bf w}_y$ be a spatial unit vector in the
direction of propagation of the wave, we may consider the set of linear
polarization tensors
\beq
{\bf e}^{+}= {\bf w}_x \otimes {\bf w}_x - {\bf w}_y \otimes {\bf w}_y\, ,
\eeq
\beq
{\bf e}^{\times}= {\bf w}_x \otimes {\bf w}_y + {\bf w}_y \otimes {\bf w}_x\, ,
\eeq
\beq
{\bf e}^{\rm x}= {\bf w}_x \otimes {\bf w}_z + {\bf w}_z \otimes {\bf w}_x\, ,
\eeq
\beq
{\bf e}^{\rm y}= {\bf w}_y \otimes {\bf w}_z + {\bf w}_z \otimes {\bf w}_y\, ,
\eeq
\beq
{\bf e}^{\rm b}= {\bf w}_x \otimes {\bf w}_x + {\bf w}_y \otimes {\bf w}_y\, ,
\eeq
\beq
{\bf e}^{\rm l}= {\bf w}_z \otimes {\bf w}_z\, .
\eeq
Then \eq{hab} implies that there exists some %\red{synchronous}
gauge in which,
in a local Lorentz frame with Cartesian coordinates along $({\bf w}_x,\,
{\bf w}_y,\, {\bf w}_z)$, 
\beq \label{eq:polarizations}
[h_{ij}] = \begin{pmatrix}
h_{\rm b} + h_+ & h_\times & h_{\rm x} \\
h_\times & h_{\rm b} - h_+ & h_{\rm y} \\
h_{\rm x} & h_{\rm y} & h_{\rm l}
\end{pmatrix} ,
\eeq
where the $h_A$'s represent the amplitudes of the linear polarizations: plus
($+$), cross ($\times$), vector x (x), vector y (y), breathing (b) and
longitudinal (l). The effect of each of these modes on a ring of
freely-falling particles is represented in \fig{circles}.

Polarizations may be characterized by their behavior under Lorentz
transformations, and different theories may be classified according to the
polarizations they allow, as seen by different observers; this is known as the
E(2) or {\em Eardley} classification \cite{Eardley1973a, Eardley1973b}. From a
field-theoretic perspective, the two tensor modes, the two vector modes and the
breathing (transverse) scalar mode correspond to the helicity $\pm2$, helicity
$\pm1$, and helicity $0$ states of a massive spin-2 particle (the graviton).
The remaining longitudinal scalar mode is usually linked to a ghost-like degree
of freedom (associated with the trace). 
% associated with the trace of the metric.
This correspondence between geometric (Eardly's classification) and
field-theoretic (Wigner's classification) language is, however, limited because
the the E(2) classification is only semi-Lorentz-invariant (although it is
usually taken to hold, at least in the weak field regime) \cite{Eardley1973b}.

Einstein's theory only allows for the existence of linear combinations of the
tensor $+$ and $\times$ polarizations \cite{tegp}. On the other hand,
scalar-tensor theories famously predict the presence of some breathing
component associated with the theory's extra scalar field \cite{Brans1961}, as
do some theories with extra dimensions \cite{Andriot2017}. On top of tensor and
scalar modes, bimetric theories, like Rosen or Lightman-Lee theories, may also
predict vector modes \cite{Lightman1973, Rosen1974, Chatziioannou2012}. The
same is true in general for massive-graviton frameworks \cite{DeRham2014}.
Furthermore, less conventional theories might, in principle, predict the
existence of vector or scalar modes \emph{only}, while still possibly being in
agreement with all other non-GW tests of GR (see e.g.~\cite{Mead2015}, for an
unconventional example).

\subsection{Antenna patterns} \label{sec:aps}

\begin{figure}
\subfloat[][Plus (+)]{\includegraphics[width=0.33\columnwidth]{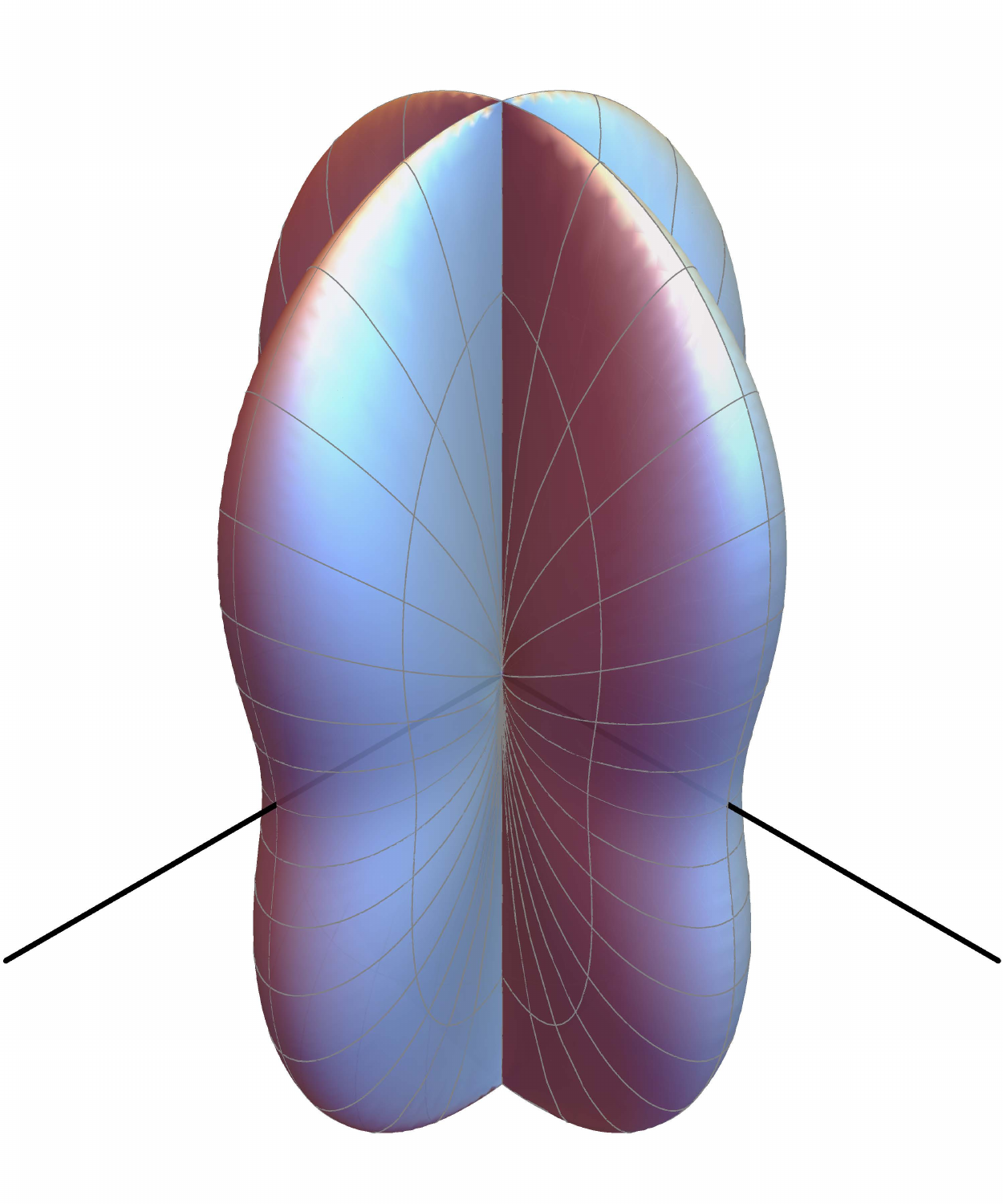}}\hspace{1cm}
\subfloat[][Cross ($\times$)]{\includegraphics[width=0.33\columnwidth]{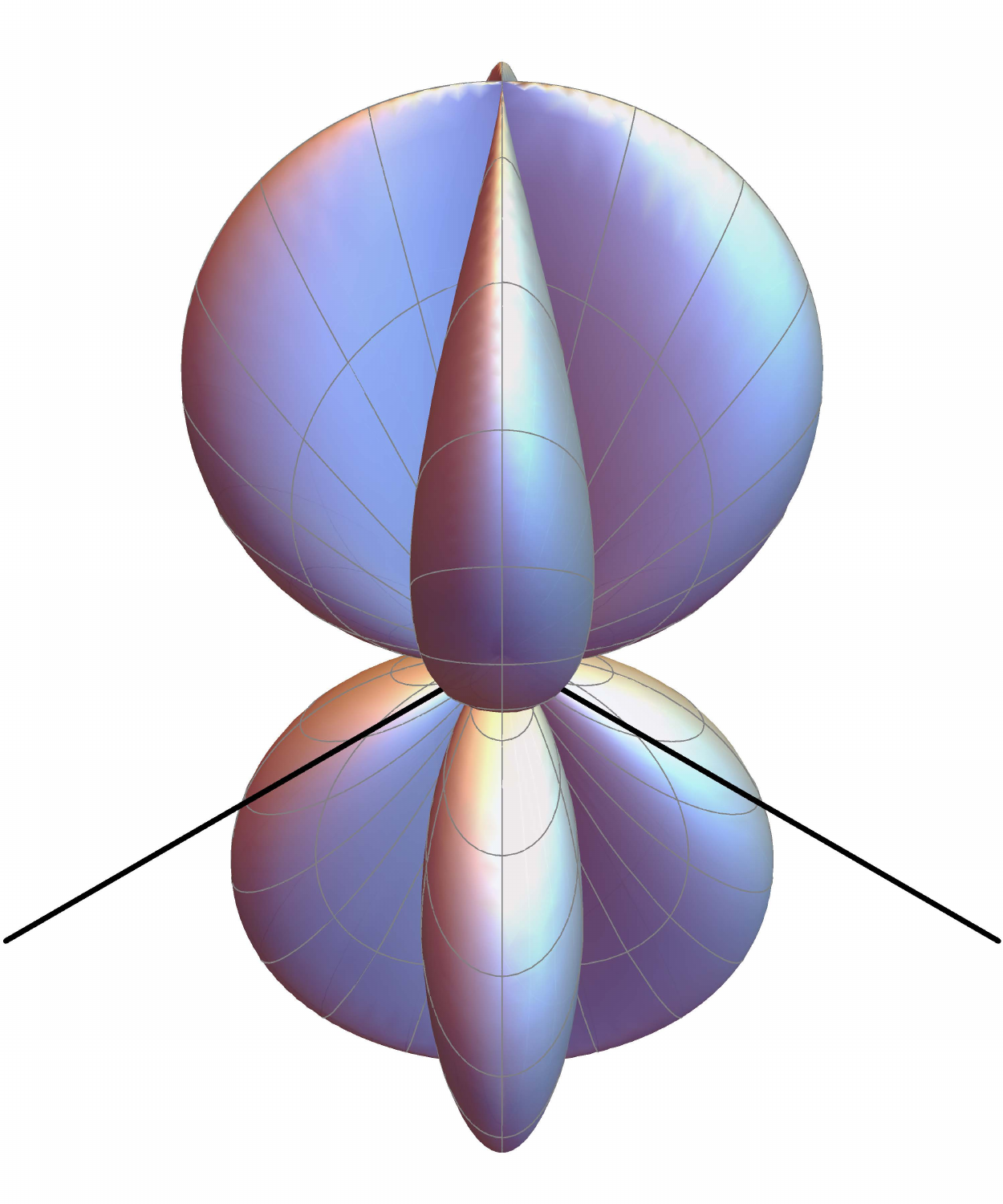}}\\
\subfloat[][Vector-x (x)]{\includegraphics[width=0.33\columnwidth]{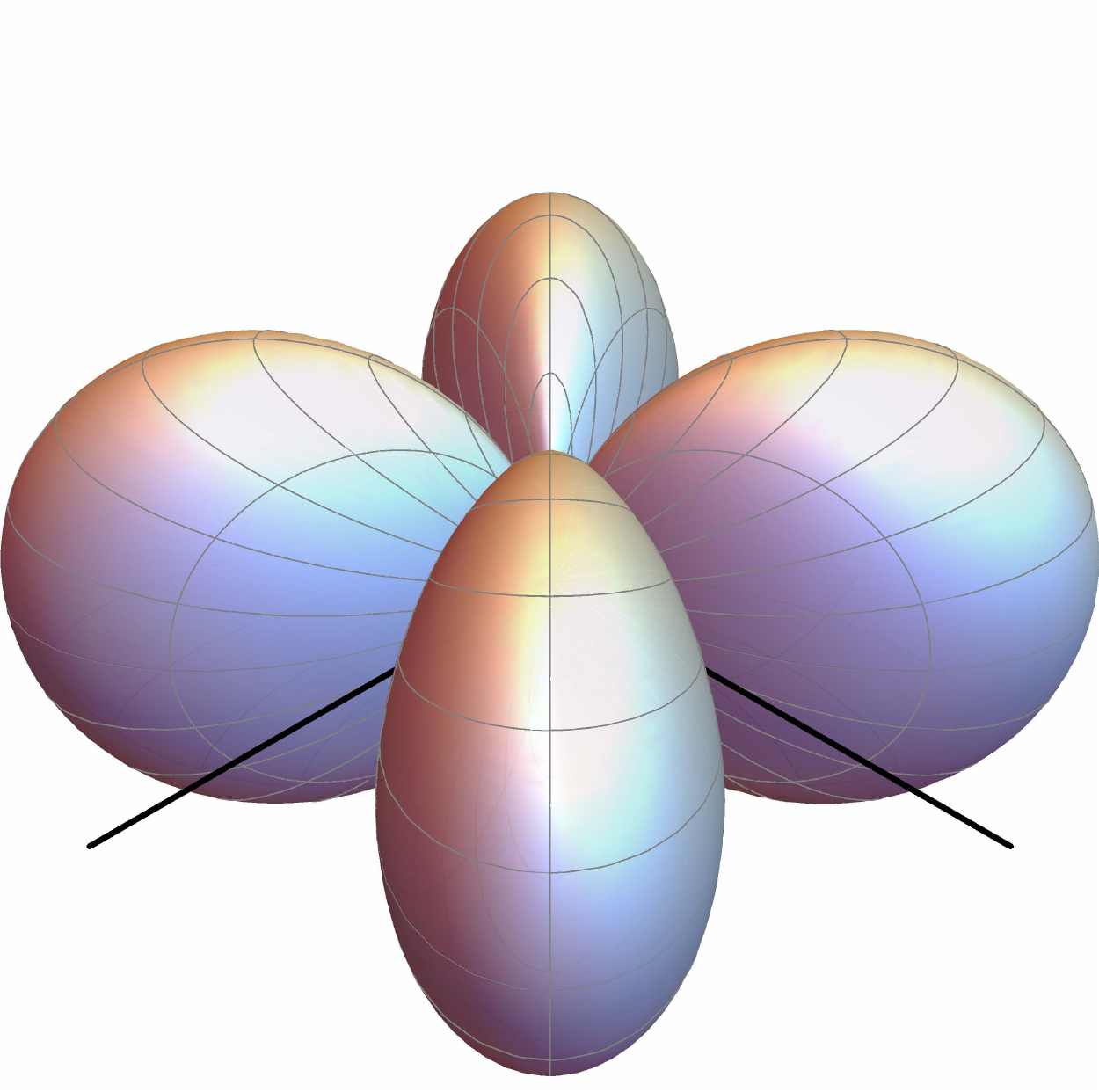}}\hspace{1cm}
\subfloat[][Vector-y (y)]{\includegraphics[width=0.33\columnwidth]{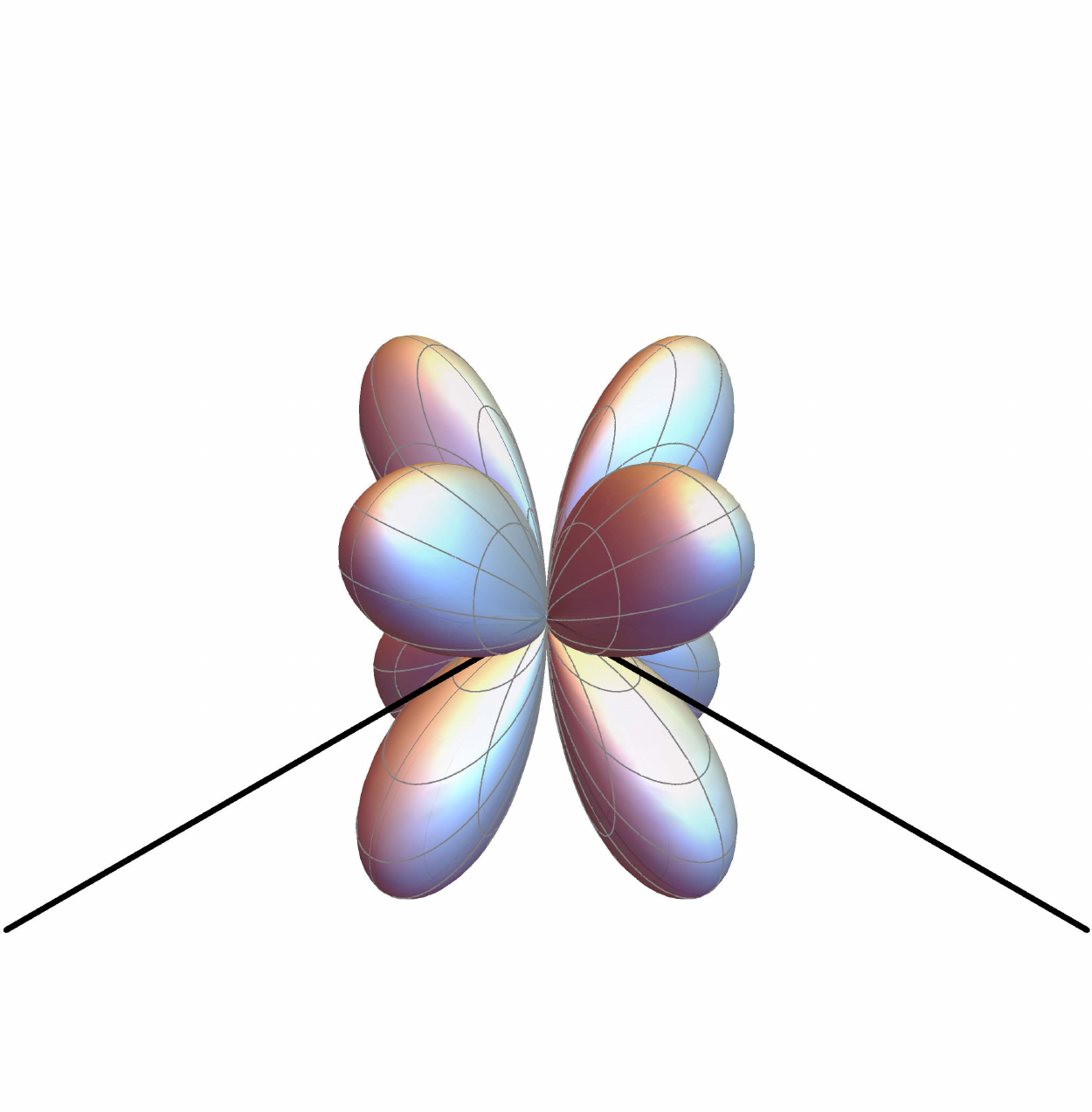}}\\
\subfloat[][Scalar (s)]{\includegraphics[width=0.33\columnwidth]{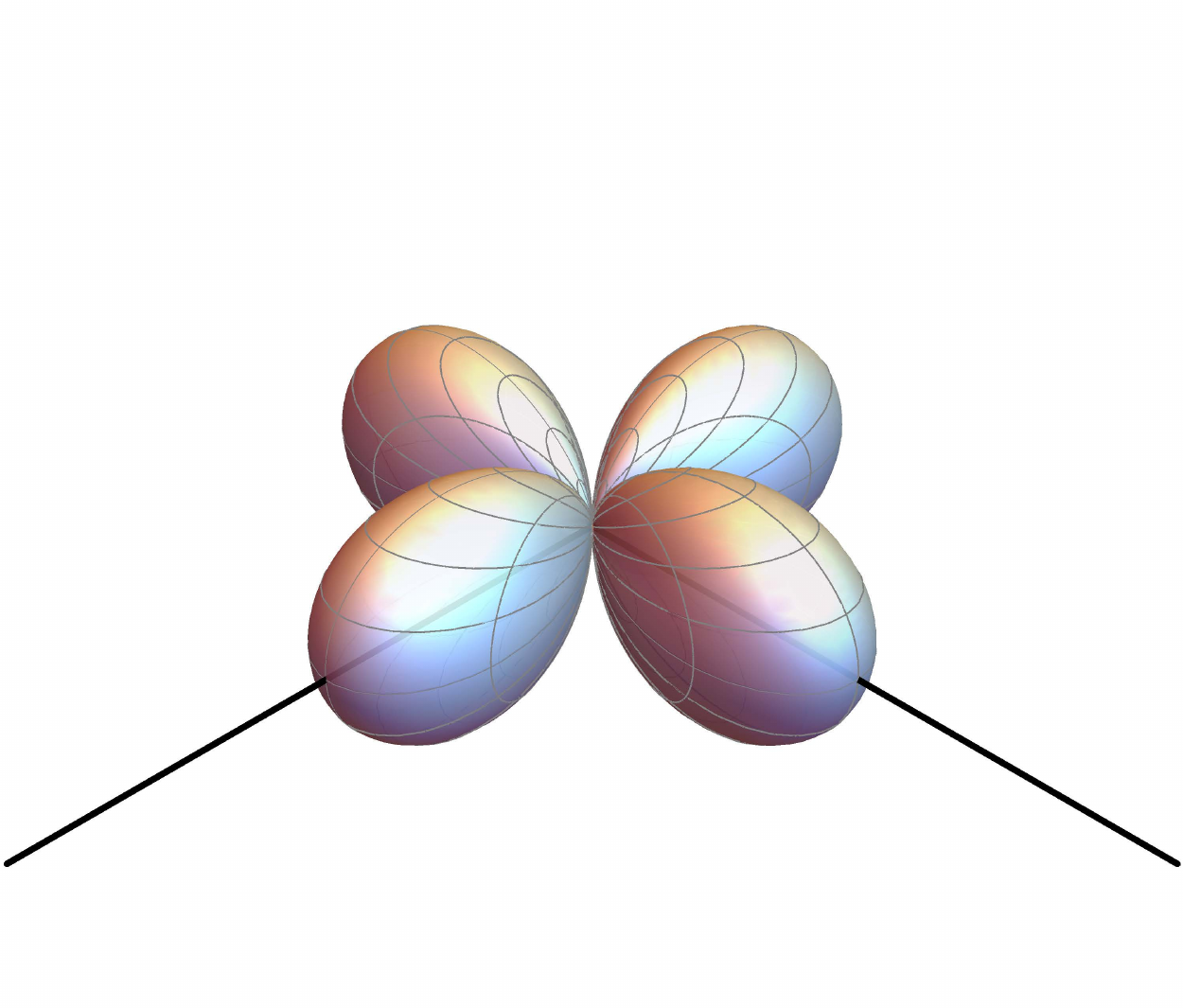}}
%\vspace{0.5in}
\caption{{\em Angular response of a quadrupolar detector to each GW
polarization}. The radial distance represents the response of a single
quadrupolar antenna to a unit-amplitude gravitational signal of a tensor (top),
vector (middle), or scalar (bottom) polarization, i.e.\ $|F_A|$ for each
polarization $A$ as given by Eqs.\ (\ref{eq:Fp_ifo}--\ref{eq:Fs_ifo}) for
$\psi=0$. The polar and azimuthal coordinates correspond to the source
location with respect to the detector, which is to be imagined as placed with
its vertex at the center of each plot and arms along the $x$ and $y$-axes. The
response is plotted to scale, such that the black lines representing the
detector arms have unit length in all plots. The response to breathing and
longitudinal modes is identical, so we only display it once and label it
``scalar''. (Reproduced from \cite{Isi2017}.)} 
\label{fig:aps}
\end{figure}

Because different polarizations have geometrically distinct effects, as
illustrated in \fig{circles}, GW detectors will react differently to each mode.
The {\em strain} produced by a GW metric perturbation $h_{ab}$ on certain
detector $I$ spatially located at ${\bf x}_I$, is given by
\beq
h_I(t) = D^{ab}_{I} h_{ab}(t, {\bf x}_I) = h_A(t, {\bf x}_I) D^{ab}_I e^A_{ab}. 
\eeq
The detector tensor, $D^{ab}$, encodes the geometry of the instrument and the
measurement it makes; for diferential-arm detectors (sometimes called {\em
quadrupolar antennas}, because of the symetries of their angular response
functions, cf.\ \fig{aps}), like LIGO and Virgo, this is
\beq
D^{ab} = \frac{1}{2}\left( d_x^{~a} d_x^{~b} - d_y^{~a} d_y^{~b} \right),
\eeq
where ${\bf d}_x$ and ${\bf d}_y$ are spatial unit vectors along the detector
arms (with common origin at the vertex ${\bf x}_I$). Although $D^{ab}$ is
technically also a function of time due to the motion of Earth with respect to
the fixed stars, in practice it can be taken as constant when treating
short-lived CBC signals, as is done here.

The $h_A(t)$'s are determined by a nontrivial combination of the source dynamics,
the details of the matter-gravity coupling, and the vacuum structure of the
theory. However, the response ({\em antenna pattern}) of detector $I$ to
polarization $A$,
\beq \label{eq:response}
F^A \equiv D^{ab}_I e^A_{ab}\, ,
\eeq
depends {\em only} on the local geometry of the gravitational wave and the
detector, irrespective of the properties of the source. This decoupling makes
the antenna patterns a unique resource for studying GW polarizations directly.

The response functions, \eq{response}, encode the effect of a linearly
$A$-polarized GW with unit amplitude, $h_A=1$. Ground-based GW detectors, like
LIGO and Virgo are quadrupolar antennas that perform low-noise measurements of
the strain associated with the differential motion of two orthogonal arms.
Their detector response functions can thus be written as \cite{Nishizawa2009,
Blaut2012, Isi2015, Poisson2014}:
\beq \label{eq:Fp}
F_{+} = \frac{1}{2} \left[ ({\bf w}_x \cdot {\bf d}_x)^2-({\bf w}_x \cdot
{\bf d}_y)^2 - ({\bf w}_y \cdot {\bf d}_x)^2+({\bf w}_y \cdot {\bf d}_y)^2
\right],
\eeq
\beq \label{eq:Fc}
F_{\times}=({\bf w}_x \cdot {\bf d}_x) ({\bf w}_y \cdot {\bf d}_x)-({\bf w}_x
\cdot {\bf d}_y) ({\bf w}_y \cdot {\bf d}_y),
\eeq
\beq \label{eq:Fx}
F_{\rm x}= ({\bf w}_x \cdot {\bf d}_x) ({\bf w}_z \cdot {\bf d}_x)- ({\bf w}_x
\cdot {\bf d}_y) ({\bf w}_z \cdot {\bf d}_y),
\eeq
\beq \label{eq:Fy}
F_{\rm y}= ({\bf w}_y \cdot {\bf d}_x) ({\bf w}_z \cdot {\bf d}_x)- ({\bf w}_y
\cdot {\bf d}_y) ({\bf w}_z \cdot {\bf d}_y),
\eeq
\beq \label{eq:Fb}
F_{\rm b}= \frac{1}{2} \left[ ({\bf w}_x \cdot {\bf d}_x)^2-({\bf w}_x \cdot
{\bf d}_y)^2+({\bf w}_y \cdot {\bf d}_x)^2-({\bf w}_y \cdot {\bf
d}_y)^2\right],
\eeq
\beq \label{eq:Fl}
F_{\rm l}=\frac{1}{2}\left[ ({\bf w}_z \cdot {\bf d}_x)^2- ({\bf w}_z
\cdot {\bf d}_y)^2 \right].
\eeq
Here, as before, the spatial vectors ${\bf d}_x$, ${\bf d}_y$ have unit norm
and point along the detector arms such that ${\bf d}_z={\bf d}_x\times{\bf
d}_y$ is the local zenith; the direction of propagation of the wave from a
source at known sky location (specified by right ascension $\alpha$, and
declination $\delta$) is given by ${\bf w}_z$, and ${\bf w}_x$, ${\bf w}_y$ are
such that ${\bf w}_z={\bf w}_x\times{\bf w}_y$.
We choose ${\bf w}_x$ to lie along the intersection of the equatorial plane of
the source with the plane of the sky, and let the angle between ${\bf w}_y$ and
the celestial north be $\psi$, the {\em polarization angle}.

Because of their symmetries, the breathing and longitudinal modes are fully
degenerate to networks of quadrupolar antennas (see e.g.\ Sec.\ VI of
\cite{Chatziioannou2012}). This means that no model-independent measurement
with such a network can possibly distinguish between the two, so it is enough
for us to consider just one of them explicitly; we will refer to the scalar
modes jointly by the subscript ``s''. (This degeneracy may not be present
for detectors with different geometries \cite{Lee2008, Chamberlin2012}.)

The response of a given differential-arm detector to signals of certain linear
polarization and direction of propagation can be written, in the local Lorentz
frame of the detector itself, as [see e.g.\ Eqs.\ (13.98) in
\cite{Poisson2014} with $\psi \rightarrow -\psi-\pi/2$, to account for the
different wave-frame definition]:
% must double check PSI dependences
\begin{align} \label{eq:Fp_ifo}
F_+(\vartheta, \varphi, \psi) = &-\frac{1}{2}\left(1+\cos^2\vartheta \right)
\cos 2\varphi \cos2\psi \nonumber\\ &-\cos\vartheta \sin2\varphi \sin2\psi~,
\end{align}
\begin{align}
F_\times(\vartheta, \varphi, \psi) &= \frac{1}{2}\left(1+\cos^2\vartheta \right)
\cos 2\varphi \sin2\psi \nonumber \\ &-\cos \vartheta \sin 2\varphi \cos2\psi~,
\end{align}
\begin{align}
F_{\rm x}(\vartheta, \varphi, \psi) &= -\sin\vartheta \sin 2\varphi \cos\psi
\nonumber\\ &+\sin\vartheta \cos\vartheta\cos 2\varphi \sin\psi~,
\end{align}
\begin{align}
F_{\rm y}(\vartheta, \varphi, \psi) &= \sin\vartheta \sin 2\varphi \sin\psi
\nonumber\\ &+\sin\vartheta \cos\vartheta
\cos 2\varphi\cos\psi~,
\end{align}
\beq \label{eq:Fs_ifo}
F_{\rm b/l}(\vartheta, \varphi, \psi) = \mp \frac{1}{2} \sin^2\vartheta
\cos 2\varphi~,
\eeq
where $\vartheta$ and $\varphi$ are the polar an azimuthal coordinates of the
source with respect to the antenna at any given time (with detector arms along
the $x$ and $y$-axes). The tensor, vector and scalar nature of the
different polarizations is evident in this form, given how each mode 
depends on $\psi$ (i.e.\ how it transforms under rotations around the 
direction of propagation).

Equations \eqref{eq:Fp_ifo}--\eqref{eq:Fs_ifo} are represented in \fig{aps} by
a spherical polar plot in which the radial coordinate corresponds to the
sensitivity given by the magnitude $|F_A|$, shown for $\psi=0$. The angular
response functions have quadrupolar symmetry around the detector's zenith,
regardless of the helicitiy of the polarization itself. This figure also makes it
clear that differential-arm detectors will generally be more sensitive to some
polarizations than others, although this will vary with the sky location of the
source. For example, for all but a few sky locations, quadrupolar antennas will
respond significantly less to a breathing signal than a plus or cross signal.

\fig{aps} shows the response of a single differential-arm detector to waves
coming from different directions in the local frame of the instrument. However,
we are usually interested in the sensitivity of a {\em network} of detectors,
and its ability to distinguish the different polarizations. To visualize this,
define the effective response to each of the helicities, for a given source
sky-location $(\alpha,\, \delta)$ and detector $I$:
\beq
|F_{\rm t}^I(\alpha, \delta)| \equiv \sqrt{F_+^I(\alpha, \delta)^2 +
F_\times^I(\alpha, \delta)^2}\, ,
\eeq
\beq
|F_{\rm v}^I(\alpha, \delta)| \equiv \sqrt{F_{\rm x}^I(\alpha, \delta)^2 + F_{\rm
y}^I(\alpha, \delta)^2}\, ,
\eeq
\begin{align}
|F_{\rm s}^I(\alpha, \delta)| &\equiv \sqrt{F_{\rm b}^I(\alpha, \delta)^2 + F_{\rm
l}^I(\alpha, \delta)^2} \\
&= \sqrt{2}\, |F^I_{\rm b}(\alpha, \delta)|\, , \nonumber
\end{align}
for tensor, vector and scalar waves respectively. (Here, since we are not
dealing with any specific source, we {\em define} our polarization frame
letting $\psi=0$.) For a network of $N$ detectors, we may then construct an
effective response vector for each of the polarization sets above,
\beq \label{eq:ap_vector}
\vec{F}_H(\alpha, \delta) \equiv \left( |F_H^1(\alpha, \delta)|,\, \dots,
|F_H^N(\alpha, \delta)| \right),
\eeq
for $H \in \{\rm t,\, v,\, s\}$. Finally, we may compare the overall sensitivity of
the network to different polarizations by defining the {\em overlap}, as a 
normalized inner product between two of these vectors.

\begin{figure}
\subfloat[][Scalar]{\includegraphics[width=\columnwidth]{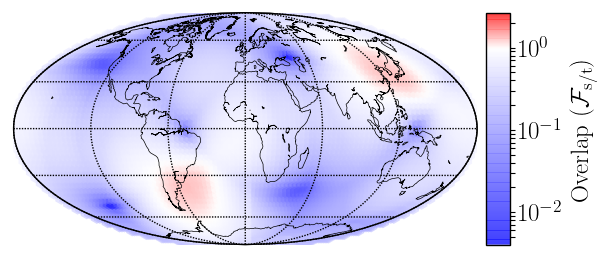}}\\
\subfloat[][Vector]{\includegraphics[width=\columnwidth]{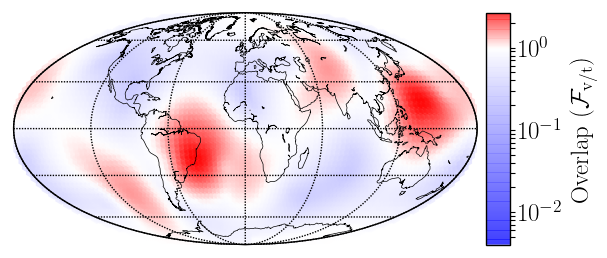}}\\
\caption{{\em Overlaps of LIGO-Virgo network effective antenna patterns.} The
normalized inner-products of \eq{ap_overlap} for the three-instrument network.
The top plot compares scalar to tensor (${\cal F}_{\rm s/t}$), and the bottom
one compares vector to tensor (${\cal F}_{\rm v/t}$). Blue (red) marks regions
for which the effective nontentor response is greater (less) than tensor. A map
of Earth is overlaid for reference.}
\label{fig:ap_skymap}
\end{figure}

For instance, to compare the effective scalar or vector network sensitivity to
the tensor one, we may look at the overlap factor:
\beq \label{eq:ap_overlap}
{\cal F}_{H/{\rm t}}(\alpha, \delta) = \frac{\vec{F}_H(\alpha, \delta) \cdot
\vec{F}_{\rm t}(\alpha, \delta)}{\vec{F}_{\rm t}(\alpha, \delta) \cdot
\vec{F}_{\rm t}(\alpha, \delta)},
\eeq
which will take values greater (less) than unity if the response to
polarizations $H$ is better (worse) than to tensor, with ${\cal F}_{\rm
t/t}(\alpha, \delta)=1$ by construction. The scalar and vector overlaps with
tensor are displayed for the LIGO-Virgo network in the skymap of
\fig{ap_skymap}, over a map of Earth for reference. Colored regions roughly
correspond to areas in the sky for which the tensor and nontensor responses of
the network are highly distinguishable. The patterns are anchored to angular
locations with respect to Earth (not the fixed stars), and is determined by the
specific location and orientation of the three detectors.

\begin{figure}
\includegraphics[width=\columnwidth]{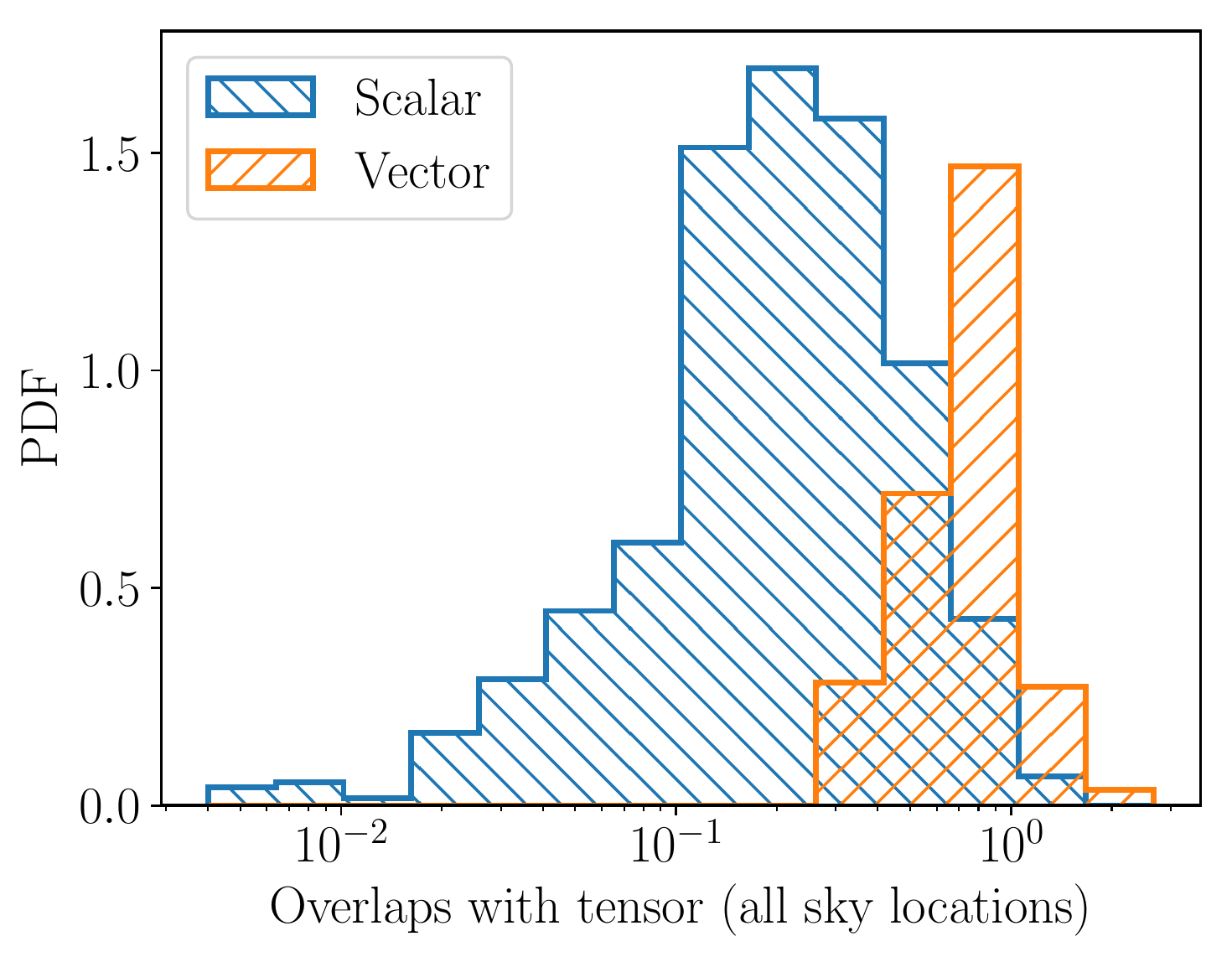}
\caption{{\em Overlaps of LIGO-Virgo network effective antenna patterns.} The
normalized inner-products of \eq{ap_overlap} for the three-instrument network.
The top plot compares scalar to tensor (${\cal F}_{\rm s/t}$), and the bottom
one compares vector to tensor (${\cal F}_{\rm v/t}$). Blue (red) marks regions
for which the effective nontentor response is greater (less) than tensor. A map
of Earth is overlaid for reference.}
\label{fig:ap_histogram}
\end{figure}

Averaged over all sky locations, the response of the network is worse for
scalar signals than tensor ones, which is apparent from the top skymap in
\fig{ap_skymap} and the distribution in \fig{ap_histogram}. This is expected
given that each interferometer is individually less sensitive to scalar waves,
as seen in \fig{aps}. On average, there is no significant difference between
vector and tensor responses.

\section{Method} \label{sec:method}

Ideally, we would like to unequivocally measure the polarizations of the GW
that produced a given transient strain signal in our detector network.
Formally, this would mean finding which of the seven possible Bayesian
hypotheses the data favor: pure tensor ($\hyp{t}$), pure vector ($\hyp{v}$),
pure scalar ($\hyp{s}$), scalar-tensor ($\hyp{st}$), vector-tensor
($\hyp{tv}$), scalar-vector ($\hyp{sv}$), or scalar-vector-tensor
($\hyp{stv}$). A comprehensive Bayesian treatment of this polarization
model-selection problem was presented in \cite{Isi2017} for the case of
continuous signals from known pulsars, later applied to stochastic GW
backgrounds in \cite{Callister2017}, and could be easily by adapted to the case
of transient signals considered here.

Yet, a simple counting argument is enough to show that three detectors are not
sufficient to break {\em all} degeneracies between the five distinguishable GW
polarizations using transient signals \cite{Chatziioannou2012, tegp}.
Therefore, with the current LIGO-Virgo network, we expect the results of an
all-encompassing model-selection analysis, as discussed above, to be
inconclusive or dominated by priors. Nevertheless, we may still attempt to
distinguish between {\em some} of the possible hypotheses.

As mentioned in the introduction, {\em all LIGO-only observations so far are
consistent with the extreme scenario of GWs being composed of purely vector or
purely scalar polarizations}.
%which is a sign of how little we have been able to learn experimentally about
%this fundamental property of GWs.
Therefore, here we will focus on the problem of directly distinguishing between
these theoretically far-fetched, yet phenomenologically valid, possibilities.
That is, we will study our ability to choose between $\hyp{s}$ vs $\hyp{t}$,
and between $\hyp{v}$ vs $\hyp{t}$. Importantly, this is qualitatively distinct
from the more standard question about the presence of small nontensorial
components in addition to the tensor wave predicted by GR. Although perhaps not
as interesting as these ``mixed'' polarization studies (which, as explained
above, will not fully succeed with current detectors), the problem of
distinguishing between the ``pure'' polarization cases is well-defined and
experimentally valuable.

We would like to ask the question: {\em is it geometrically possible that a
\underline{given} strain signal observed in the LIGO-Virgo network was produced
by a GW with polarization other than GR's tensor $+$ and $\times$?}
The only way for us to answer this question is to probe the
antenna patterns of our instruments, \eq{response}, which are a direct
manifestation of local geometry only (polarizations and detector geometry),
independent of source or the details of the underlying theory (see Sec.\
\ref{sec:aps}). We may thus exploit the difference in the response of the
network to the different polarizations (\fig{ap_skymap}). 

One way to extract polarization information using the antenna patterns would be
to construct linear combinations of the detector outputs that are guaranteed to
contain no tensorial signal \cite{Chatziioannou2012}. If coherent power (as
seen by, e.g.\ a wavelet analysis) remains in such a {\em null-stream}, then
that signal could not have been produced by a tensor (GR) wave. This approach
has the strong advantage that it requires no knowledge of the spectral features
of the signal whatsoever. However, to construct null-streams one needs to very
accurately know the location of the source {\em a priori}, which is never the
case without an electromagnetic counterpart (or more detectors). 

Alternatively, one could carry out a morphology-independent sine-Gaussian
analysis (e.g.~using \texttt{BayesWave} \cite{Cornish2014, Littenberg2016}) to
reconstruct the best-fit unmodeled waveform from the data, and use that to
extract information about times of arrival, phase offsets and relative
amplitudes at different detectors. One could then just replace the tensor
antenna patterns used in the signal reconstruction by their scalar or vector
counterparts, and see how well each case fits the data (as measured by a Bayes
factor). In such test, {\em no polarization information is extracted from the
phase evolution}. In particular, the waveform reconstruction is only used to
infer the source location from the time lag between detectors, and the
best-fitting combination of antenna patterns from the amplitudes and phases at
peak energy.  (See pedagogical example in Sec.\ \ref{sec:example} below.) An
analysis like this was implemented for scalar modes and applied to the GW150914
signal, yielding no conclusive results as mentioned above \cite{gw150914_tgr}.

However, all signals observed by LIGO so far are exceptionally well described
by GR CBC waveforms \cite{o1bbh, gw150914_tgr, gw151226, gw170104}. This match
is established on a case-by-case basis through comparisons between the GR
templates and morphology independent burst reconstructions of the signal in the
data, and is largely independent of the polarization. In fact, for any of these
confident detections, the waveform reconstructed from burst analyses is
effectively identical to a GR template. As emphasized above, in the
pure-polarization test ($\hyp{s}$ vs $\hyp{t}$, or $\hyp{v}$ vs $\hyp{t}$) all
that matters is that most of the signal power is captured by the template,
regardless of small potential mismatches in the phasing. Therefore, we may
carry out the same study proposed in the previous paragraph using GR waveforms
to fit the data, while replacing the tensor antenna patterns with those of
different polarizations.

In other words, when the signal is clearly well-captured by a GR template, we
may use that directly to extract polarization information from the antenna
patterns in a model independent way, without implicitly assuming that the GW
that caused it was tensor polarized as GR predicts. The waveform reconstruction
will be dominated by the measurement at the most sensitive detector, while the
amplitude information is encoded in the relations between measurements by
different detectors.

Whether we use GR templates or a collection of sine-Gaussians to reconstruct
the waveform, the effect of changing the antenna patterns will always result in
different inferred sky location and orientation for the source. Yet, not all
antenna patterns will be equally consistent with the observed relative
amplitudes, phase offsets and delays between the signals in our three
detectors---this will result in a poorer signal likelihood, and hence odds
favoring tensor vs nontensor. Precisely because the waveform used to capture
the signal is the same, we know that any difference between the tensor and
nontensor results {\em must} come from the antenna patterns (polarizations).

\blue{This approach does not extract any information from the specific phase
evolution of the signal}, and is insensitive to small changes in the waveform.
Therefore, using a GR template to measure the signal power is justified, and does
not imply a contradiction when testing for nontensorial polarizations. For
the purpose of this study, the CBC signal is just probing the impulse response
function of our network, and the same results would be obtained if the waveform
was just a Delta function rather than a chirp.

\subsection{Toy example} \label{sec:example}

For concreteness, consider the example of an elliptically-polarized,
two-component GW (e.g.\ two tensor modes, or two vector modes) with waveform
roughly described by a simple sine-Gaussian wavepacket, with some
characteristic frequency $\Omega$ and relaxation time $\tau$. Letting $t$ be
the time measured at Earth's center, then the strain measured by a given
detector $I$ will be:
\beq \label{eq:toy_h}
h_I(t) = \Re\left[ A \left( F_{1}^I + i \epsilon F_{2}^I \right)
e^{i\Omega (t-t_0-\delta t_I)} \right] e^{-(t-t_0-\delta t_I)^2/\tau^2},
\eeq
where $F_1^I$ and $F_2^I$ are the responses of detector $I$ to the two
polarizations, $A\equiv|A|e^{i\phi_0}$ is a complex-valued amplitude,
$\epsilon$ is an ellipticity parameter controlling the relative amounts of each
polarization, and $\Re$ denotes the real part. Also, $t_0$ marks the time of
arrival at Earth's center, which is delayed with respect to each interferometer
by
\beq
\delta t_I = \hat{{\bf n}} \cdot {\bf x}_I / c\, ,
\eeq
where $\hat{{\bf n}}$ is a unit vector from Earth to the source, and ${\bf
x}_I$ joins Earth's center to the detector (with magnitude equal to Earth's
radius). Here we are assuming that the GW travels at the speed of light, $c$.

The signal of \eq{toy_h} may be written more simply as
\beq
h_I(t) = {\cal A}_I \cos [\Omega (t - \Delta t_I) + \Phi_I] \, e^{-(t-\Delta
t_I)^2/\tau^2}\, ,
\eeq
after defining the three main observables at each detector:
\beq
{\cal A}_I \equiv |A| \left| F_{1}^I + i \epsilon F_{2}^I \right|\, ,
\eeq
\beq
\Phi_I \equiv \phi_0 + \arctan (\epsilon F_2^I/F^I_1)\, ,
\eeq
\beq
\Delta t_I \equiv t_0 + \delta t_I 
\eeq
From the output of three detectors ($H$, $L$, $V$), we may implement a simple
inference analysis to extract these three numbers for the signal as seen by
each instrument. The times at peak amplitude provide the three $\Delta t_I$'s,
while measurements of the phase and amplitude at peak itself give the $\Phi$'s
and ${\cal A}_I$'s respectively. As always, recovery of all these parameters
will be negatively affected by instrumental noise.

The three timing measurements alone suffice to recover the sky location of the
source, $\hat{\bf n}$. With this knowledge, it is then possible to compute the
values of all the corresponding antenna response functions, and thus obtain
predictions for the $(F_1^I + i \epsilon F_2^I)$ factors for any given
ellipticity. Ratios of amplitudes and phase differences between detectors may
then be used to infer measured values for these quantities, and then find the
best fitting polarization model. This may be achieved, for instance, via a
maximum-likelihood analysis, effectively minimizing the distance between
vectors like those of \eq{ap_vector} and a similar one inferred from the
data. (Note $|A|$, $\phi_0$, and $\epsilon$ are nuisance parameters, and can be
marginalized over.)

Although for this example we used a simple sine-Gaussian wavepacket to measure the
signal, at no point we made use of the specific details of this phase evolution.
The only requirement is that the GW have a well-defined peak, in order to extract
meaningful information about how the relative timing, phase and amplitude of
this peak as seen by different detectors. In particular, this analysis would work
precisely the same way if CBC-like chirp waveform was used, as long as most of the
power in the actual signal is indeed captured by such a template.

This toy analysis makes the dependence on $A_I$, $\Phi_I$, $\Delta t_I$
explicit. In reality, when studying actual data, one would ideally implement a
full Bayesian analysis, marginalizing over all parameters to compute evidences
for the different polarization hypotheses ($\hyp{t}$, $\hyp{v}$, $\hyp{s}$),
and to produce the Bayes factors (likelihood ratios) of interest. This can be
achieved using a code like \texttt{LALInference} \cite{Veitch2015}. The
polarization information extracted by this more rigorous analysis would still,
nonetheless, effectively come from the values of $A_I$, $\Phi_I$, $\Delta t_I$.
As emphasized before, this is the case whether one uses GR templates or a
collection of sine-Gaussians to capture the signal power.

\section{Conclusion}

By extracting polarization information from the antenna patterns we may
directly probe the geometry of the GW metric perturbation (i.e.~the directions
along which space is stretched and squeezed by the passing wave) from its
projection onto our detector network. With transient signals, instruments at five
or more different orientations would be needed to break all degeneracies
between the five independent (as seen by differential-arm detectors)
polarizations allowed by generic metric theories of gravity. However, we may
already distinguish between some of the possibilities using the current
LIGO-Virgo network. How well we can do this will depend on the specific
properties of each transient event (mainly, sky location).

The kind of geometric observational statement discussed in this note is
independent of any theory or source model, and is only possible with the
addition of Virgo to the network. Although here we focused on the problem of
distinguishing between ``pure'' polarization states (tensor, vector or
scalar), the case of ``mixed'' polarizations will be addressed in future work.
More details and a demonstration of the analysis proposed here on simulated
signals will be provided soon in an expanded version of this document.

\begin{acknowledgments}
% Personal
% LIGO
LIGO was constructed by the California Institute of Technology and
Massachusetts Institute of Technology with funding from the National Science
Foundation and operates under cooperative agreement PHY-0757058.
% ARCCA
% We are grateful for computational resources provided by Cardiff University, and
% funded by an STFC grant supporting UK Involvement in the Operation of Advanced
% LIGO.
% Matplotlib
% Plots produced using \texttt{Matplotlib} \cite{Hunter2007}.
% DCC
This paper carries LIGO Document Number \dcc{}.
\end{acknowledgments}

\bibliography{gw,gr,statistics,hep}

\begin{thebibliography}{39}%
\makeatletter
\providecommand \@ifxundefined [1]{%
 \@ifx{#1\undefined}
}%
\providecommand \@ifnum [1]{%
 \ifnum #1\expandafter \@firstoftwo
 \else \expandafter \@secondoftwo
 \fi
}%
\providecommand \@ifx [1]{%
 \ifx #1\expandafter \@firstoftwo
 \else \expandafter \@secondoftwo
 \fi
}%
\providecommand \natexlab [1]{#1}%
\providecommand \enquote  [1]{``#1''}%
\providecommand \bibnamefont  [1]{#1}%
\providecommand \bibfnamefont [1]{#1}%
\providecommand \citenamefont [1]{#1}%
\providecommand \href@noop [0]{\@secondoftwo}%
\providecommand \href [0]{\begingroup \@sanitize@url \@href}%
\providecommand \@href[1]{\@@startlink{#1}\@@href}%
\providecommand \@@href[1]{\endgroup#1\@@endlink}%
\providecommand \@sanitize@url [0]{\catcode `\\12\catcode `\$12\catcode
  `\&12\catcode `\#12\catcode `\^12\catcode `\_12\catcode `\%12\relax}%
\providecommand \@@startlink[1]{}%
\providecommand \@@endlink[0]{}%
\providecommand \url  [0]{\begingroup\@sanitize@url \@url }%
\providecommand \@url [1]{\endgroup\@href {#1}{\urlprefix }}%
\providecommand \urlprefix  [0]{URL }%
\providecommand \Eprint [0]{\href }%
\providecommand \doibase [0]{http://dx.doi.org/}%
\providecommand \selectlanguage [0]{\@gobble}%
\providecommand \bibinfo  [0]{\@secondoftwo}%
\providecommand \bibfield  [0]{\@secondoftwo}%
\providecommand \translation [1]{[#1]}%
\providecommand \BibitemOpen [0]{}%
\providecommand \bibitemStop [0]{}%
\providecommand \bibitemNoStop [0]{.\EOS\space}%
\providecommand \EOS [0]{\spacefactor3000\relax}%
\providecommand \BibitemShut  [1]{\csname bibitem#1\endcsname}%
\let\auto@bib@innerbib\@empty
%</preamble>
\bibitem [{\citenamefont {Abbott~et al.}\ \emph
  {et~al.}(2016{\natexlab{a}})\citenamefont {Abbott~et al.}, \citenamefont
  {{(The LIGO Scientific Collaboration}},\ and\ \citenamefont {{The Virgo
  Collaboration)}}}]{gw150914}%
  \BibitemOpen
  \bibfield  {author} {\bibinfo {author} {\bibfnamefont {B.~P.}\ \bibnamefont
  {Abbott~et al.}}, \bibinfo {author} {\bibnamefont {{(The LIGO Scientific
  Collaboration}}}, \ and\ \bibinfo {author} {\bibnamefont {{The Virgo
  Collaboration)}}},\ }\href {\doibase 10.1103/PhysRevLett.116.061102}
  {\bibfield  {journal} {\bibinfo  {journal} {Phys. Rev. Lett.}\ }\textbf
  {\bibinfo {volume} {116}},\ \bibinfo {pages} {061102} (\bibinfo {year}
  {2016}{\natexlab{a}})}\BibitemShut {NoStop}%
\bibitem [{\citenamefont {Abbott~et al.}\ \emph
  {et~al.}(2016{\natexlab{b}})\citenamefont {Abbott~et al.}, \citenamefont
  {{(The LIGO Scientific Collaboration}},\ and\ \citenamefont {{The Virgo
  Collaboration)}}}]{gw151226}%
  \BibitemOpen
  \bibfield  {author} {\bibinfo {author} {\bibfnamefont {B.~P.}\ \bibnamefont
  {Abbott~et al.}}, \bibinfo {author} {\bibnamefont {{(The LIGO Scientific
  Collaboration}}}, \ and\ \bibinfo {author} {\bibnamefont {{The Virgo
  Collaboration)}}},\ }\href {\doibase 10.1103/PhysRevLett.116.241103}
  {\bibfield  {journal} {\bibinfo  {journal} {Phys. Rev. Lett.}\ }\textbf
  {\bibinfo {volume} {116}},\ \bibinfo {pages} {241103} (\bibinfo {year}
  {2016}{\natexlab{b}})}\BibitemShut {NoStop}%
\bibitem [{\citenamefont {{(The LIGO Scientific Collaboration}}\ and\
  \citenamefont {{The Virgo Collaboration)}}(2016)}]{o1bbh}%
  \BibitemOpen
  \bibfield  {author} {\bibinfo {author} {\bibnamefont {{(The LIGO Scientific
  Collaboration}}}\ and\ \bibinfo {author} {\bibnamefont {{The Virgo
  Collaboration)}}},\ }\href {\doibase 10.1103/PhysRevX.6.041015} {\bibfield
  {journal} {\bibinfo  {journal} {Phys. Rev. X}\ }\textbf {\bibinfo {volume}
  {6}},\ \bibinfo {pages} {041015} (\bibinfo {year} {2016})},\ \Eprint
  {http://arxiv.org/abs/1606.04856} {arXiv:1606.04856} \BibitemShut {NoStop}%
\bibitem [{\citenamefont {Abbott}\ \emph {et~al.}(2017)\citenamefont {Abbott},
  \citenamefont {{(The LIGO Scientific Collaboration}},\ and\ \citenamefont
  {{The Virgo Collaboration)}}}]{gw170104}%
  \BibitemOpen
  \bibfield  {author} {\bibinfo {author} {\bibfnamefont {B.~P.}\ \bibnamefont
  {Abbott}}, \bibinfo {author} {\bibnamefont {{(The LIGO Scientific
  Collaboration}}}, \ and\ \bibinfo {author} {\bibnamefont {{The Virgo
  Collaboration)}}},\ }\href {\doibase 10.1103/PhysRevLett.118.221101}
  {\bibfield  {journal} {\bibinfo  {journal} {Phys. Rev. Lett.}\ }\textbf
  {\bibinfo {volume} {118}},\ \bibinfo {pages} {221101} (\bibinfo {year}
  {2017})},\ \Eprint {http://arxiv.org/abs/1706.01812} {arXiv:1706.01812}
  \BibitemShut {NoStop}%
\bibitem [{\citenamefont {Abbott~et al.}\ \emph
  {et~al.}(2016{\natexlab{c}})\citenamefont {Abbott~et al.}, \citenamefont
  {{(The LIGO Scientific Collaboration}},\ and\ \citenamefont {{The Virgo
  Collaboration)}}}]{gw150914_tgr}%
  \BibitemOpen
  \bibfield  {author} {\bibinfo {author} {\bibfnamefont {B.~P.}\ \bibnamefont
  {Abbott~et al.}}, \bibinfo {author} {\bibnamefont {{(The LIGO Scientific
  Collaboration}}}, \ and\ \bibinfo {author} {\bibnamefont {{The Virgo
  Collaboration)}}},\ }\href {\doibase 10.1103/PhysRevLett.116.221101}
  {\bibfield  {journal} {\bibinfo  {journal} {Phys. Rev. Lett.}\ }\textbf
  {\bibinfo {volume} {116}},\ \bibinfo {pages} {221101} (\bibinfo {year}
  {2016}{\natexlab{c}})},\ \Eprint {http://arxiv.org/abs/1602.03841}
  {arXiv:1602.03841} \BibitemShut {NoStop}%
\bibitem [{\citenamefont {Will}(1993)}]{tegp}%
  \BibitemOpen
  \bibfield  {author} {\bibinfo {author} {\bibfnamefont {C.~M.}\ \bibnamefont
  {Will}},\ }\href@noop {} {\emph {\bibinfo {title} {{Theory and experiment in
  gravitational physics}}}},\ \bibinfo {edition} {revised ed}\ ed.\ (\bibinfo
  {publisher} {Cambridge University Press},\ \bibinfo {address} {Cambridge},\
  \bibinfo {year} {1993})\BibitemShut {NoStop}%
\bibitem [{\citenamefont {Will}(2014)}]{Will2006}%
  \BibitemOpen
  \bibfield  {author} {\bibinfo {author} {\bibfnamefont {C.~M.}\ \bibnamefont
  {Will}},\ }\href {\doibase 10.12942/lrr-2014-4} {\bibfield  {journal}
  {\bibinfo  {journal} {Living Rev. Relativ.}\ }\textbf {\bibinfo {volume}
  {17}} (\bibinfo {year} {2014}),\ 10.12942/lrr-2014-4}\BibitemShut {NoStop}%
\bibitem [{\citenamefont {Chatziioannou}\ \emph {et~al.}(2012)\citenamefont
  {Chatziioannou}, \citenamefont {Yunes},\ and\ \citenamefont
  {Cornish}}]{Chatziioannou2012}%
  \BibitemOpen
  \bibfield  {author} {\bibinfo {author} {\bibfnamefont {K.}~\bibnamefont
  {Chatziioannou}}, \bibinfo {author} {\bibfnamefont {N.~N.}\ \bibnamefont
  {Yunes}}, \ and\ \bibinfo {author} {\bibfnamefont {N.}~\bibnamefont
  {Cornish}},\ }\href {\doibase 10.1103/PhysRevD.86.022004} {\bibfield
  {journal} {\bibinfo  {journal} {Phys. Rev. D}\ }\textbf {\bibinfo {volume}
  {86}},\ \bibinfo {pages} {022004} (\bibinfo {year} {2012})},\ \Eprint
  {http://arxiv.org/abs/1204.2585} {arXiv:1204.2585} \BibitemShut {NoStop}%
\bibitem [{\citenamefont {Weisberg}\ \emph {et~al.}(2010)\citenamefont
  {Weisberg}, \citenamefont {Nice},\ and\ \citenamefont
  {Taylor}}]{Weisberg2010}%
  \BibitemOpen
  \bibfield  {author} {\bibinfo {author} {\bibfnamefont {J.~M.}\ \bibnamefont
  {Weisberg}}, \bibinfo {author} {\bibfnamefont {D.~J.}\ \bibnamefont {Nice}},
  \ and\ \bibinfo {author} {\bibfnamefont {J.~H.}\ \bibnamefont {Taylor}},\
  }\href {\doibase 10.1088/0004-637X/722/2/1030} {\bibfield  {journal}
  {\bibinfo  {journal} {Astrophys. J.}\ }\textbf {\bibinfo {volume} {722}},\
  \bibinfo {pages} {1030} (\bibinfo {year} {2010})},\ \Eprint
  {http://arxiv.org/abs/1011.0718} {arXiv:1011.0718} \BibitemShut {NoStop}%
\bibitem [{\citenamefont {Freire}\ \emph {et~al.}(2012)\citenamefont {Freire},
  \citenamefont {Wex}, \citenamefont {Esposito-Far{\`{e}}se}, \citenamefont
  {Verbiest}, \citenamefont {Bailes}, \citenamefont {Jacoby}, \citenamefont
  {Kramer}, \citenamefont {Stairs}, \citenamefont {Antoniadis},\ and\
  \citenamefont {Janssen}}]{Freire2012}%
  \BibitemOpen
  \bibfield  {author} {\bibinfo {author} {\bibfnamefont {P.~C.~C.}\
  \bibnamefont {Freire}}, \bibinfo {author} {\bibfnamefont {N.}~\bibnamefont
  {Wex}}, \bibinfo {author} {\bibfnamefont {G.}~\bibnamefont
  {Esposito-Far{\`{e}}se}}, \bibinfo {author} {\bibfnamefont {J.~P.~W.}\
  \bibnamefont {Verbiest}}, \bibinfo {author} {\bibfnamefont {M.}~\bibnamefont
  {Bailes}}, \bibinfo {author} {\bibfnamefont {B.~A.}\ \bibnamefont {Jacoby}},
  \bibinfo {author} {\bibfnamefont {M.}~\bibnamefont {Kramer}}, \bibinfo
  {author} {\bibfnamefont {I.~H.}\ \bibnamefont {Stairs}}, \bibinfo {author}
  {\bibfnamefont {J.}~\bibnamefont {Antoniadis}}, \ and\ \bibinfo {author}
  {\bibfnamefont {G.~H.}\ \bibnamefont {Janssen}},\ }\href {\doibase
  10.1111/j.1365-2966.2012.21253.x} {\bibfield  {journal} {\bibinfo  {journal}
  {Mon. Not. R. Astron. Soc.}\ }\textbf {\bibinfo {volume} {423}},\ \bibinfo
  {pages} {3328} (\bibinfo {year} {2012})}\BibitemShut {NoStop}%
\bibitem [{\citenamefont {Stairs}(2003)}]{Stairs2003}%
  \BibitemOpen
  \bibfield  {author} {\bibinfo {author} {\bibfnamefont {I.~H.}\ \bibnamefont
  {Stairs}},\ }\href {\doibase 10.12942/lrr-2003-5} {\bibfield  {journal}
  {\bibinfo  {journal} {Living Rev. Relativ.}\ }\textbf {\bibinfo {volume}
  {6}},\ \bibinfo {pages} {5} (\bibinfo {year} {2003})}\BibitemShut {NoStop}%
\bibitem [{\citenamefont {Wex}(2014)}]{Wex2014}%
  \BibitemOpen
  \bibfield  {author} {\bibinfo {author} {\bibfnamefont {N.}~\bibnamefont
  {Wex}},\ }\href {http://arxiv.org/abs/1402.5594} {\enquote {\bibinfo {title}
  {{Testing Relativistic Gravity with Radio Pulsars}},}\ } (\bibinfo {year}
  {2014}),\ \Eprint {http://arxiv.org/abs/1402.5594} {arXiv:1402.5594}
  \BibitemShut {NoStop}%
\bibitem [{\citenamefont {Berezhiani}\ \emph {et~al.}(2007)\citenamefont
  {Berezhiani}, \citenamefont {Comelli}, \citenamefont {Nesti},\ and\
  \citenamefont {Pilo}}]{Berezhiani2007}%
  \BibitemOpen
  \bibfield  {author} {\bibinfo {author} {\bibfnamefont {Z.}~\bibnamefont
  {Berezhiani}}, \bibinfo {author} {\bibfnamefont {D.}~\bibnamefont {Comelli}},
  \bibinfo {author} {\bibfnamefont {F.}~\bibnamefont {Nesti}}, \ and\ \bibinfo
  {author} {\bibfnamefont {L.}~\bibnamefont {Pilo}},\ }\href {\doibase
  10.1103/PhysRevLett.99.131101} {\bibfield  {journal} {\bibinfo  {journal}
  {Phys. Rev. Lett.}\ }\textbf {\bibinfo {volume} {99}},\ \bibinfo {pages}
  {131101} (\bibinfo {year} {2007})},\ \Eprint {http://arxiv.org/abs/0703264}
  {arXiv:0703264 [hep-th]} \BibitemShut {NoStop}%
\bibitem [{\citenamefont {Hassan}\ \emph {et~al.}(2013)\citenamefont {Hassan},
  \citenamefont {Schmidt-May},\ and\ \citenamefont {von Strauss}}]{Hassan2013}%
  \BibitemOpen
  \bibfield  {author} {\bibinfo {author} {\bibfnamefont {S.}~\bibnamefont
  {Hassan}}, \bibinfo {author} {\bibfnamefont {A.}~\bibnamefont {Schmidt-May}},
  \ and\ \bibinfo {author} {\bibfnamefont {M.}~\bibnamefont {von Strauss}},\
  }\href {\doibase 10.1007/JHEP05(2013)086} {\bibfield  {journal} {\bibinfo
  {journal} {J. High Energy Phys.}\ }\textbf {\bibinfo {volume} {2013}},\
  \bibinfo {pages} {86} (\bibinfo {year} {2013})},\ \Eprint
  {http://arxiv.org/abs/1208.1515} {arXiv:1208.1515} \BibitemShut {NoStop}%
\bibitem [{\citenamefont {Max}\ \emph {et~al.}(2017)\citenamefont {Max},
  \citenamefont {Platscher},\ and\ \citenamefont {Smirnov}}]{Max2017}%
  \BibitemOpen
  \bibfield  {author} {\bibinfo {author} {\bibfnamefont {K.}~\bibnamefont
  {Max}}, \bibinfo {author} {\bibfnamefont {M.}~\bibnamefont {Platscher}}, \
  and\ \bibinfo {author} {\bibfnamefont {J.}~\bibnamefont {Smirnov}},\ }\href
  {\doibase 10.1103/PhysRevLett.119.111101} {\bibfield  {journal} {\bibinfo
  {journal} {Phys. Rev. Lett.}\ }\textbf {\bibinfo {volume} {119}},\ \bibinfo
  {pages} {111101} (\bibinfo {year} {2017})},\ \Eprint
  {http://arxiv.org/abs/1703.07785} {arXiv:1703.07785} \BibitemShut {NoStop}%
\bibitem [{\citenamefont {Brax}\ \emph {et~al.}(2017)\citenamefont {Brax},
  \citenamefont {Davis},\ and\ \citenamefont {Noller}}]{Brax2017}%
  \BibitemOpen
  \bibfield  {author} {\bibinfo {author} {\bibfnamefont {P.}~\bibnamefont
  {Brax}}, \bibinfo {author} {\bibfnamefont {A.-C.}\ \bibnamefont {Davis}}, \
  and\ \bibinfo {author} {\bibfnamefont {J.}~\bibnamefont {Noller}},\ }\href
  {\doibase 10.1103/PhysRevD.96.023518} {\bibfield  {journal} {\bibinfo
  {journal} {Phys. Rev. D}\ }\textbf {\bibinfo {volume} {96}},\ \bibinfo
  {pages} {023518} (\bibinfo {year} {2017})},\ \Eprint
  {http://arxiv.org/abs/1703.08016} {arXiv:1703.08016} \BibitemShut {NoStop}%
\bibitem [{\citenamefont {Pontecorvo}(1957)}]{Pontecorvo1957}%
  \BibitemOpen
  \bibfield  {author} {\bibinfo {author} {\bibfnamefont {B.}~\bibnamefont
  {Pontecorvo}},\ }\href@noop {} {\bibfield  {journal} {\bibinfo  {journal}
  {Sov. Phys. JETP}\ }\textbf {\bibinfo {volume} {6}},\ \bibinfo {pages} {429}
  (\bibinfo {year} {1957})}\BibitemShut {NoStop}%
\bibitem [{\citenamefont {Pontecorvo}(1968)}]{Pontecorvo1967}%
  \BibitemOpen
  \bibfield  {author} {\bibinfo {author} {\bibfnamefont {B.}~\bibnamefont
  {Pontecorvo}},\ }\href@noop {} {\bibfield  {journal} {\bibinfo  {journal}
  {Sov. Phys. JETP}\ }\textbf {\bibinfo {volume} {26}},\ \bibinfo {pages} {984}
  (\bibinfo {year} {1968})}\BibitemShut {NoStop}%
\bibitem [{\citenamefont {Alexander}\ and\ \citenamefont
  {Yunes}(2009)}]{Alexander2009}%
  \BibitemOpen
  \bibfield  {author} {\bibinfo {author} {\bibfnamefont {S.}~\bibnamefont
  {Alexander}}\ and\ \bibinfo {author} {\bibfnamefont {N.}~\bibnamefont
  {Yunes}},\ }\href {\doibase 10.1016/j.physrep.2009.07.002} {\bibfield
  {journal} {\bibinfo  {journal} {Phys. Rep.}\ }\textbf {\bibinfo {volume}
  {480}},\ \bibinfo {pages} {1} (\bibinfo {year} {2009})},\ \Eprint
  {http://arxiv.org/abs/0907.2562} {arXiv:0907.2562} \BibitemShut {NoStop}%
\bibitem [{\citenamefont {Eardley}\ \emph
  {et~al.}(1973{\natexlab{a}})\citenamefont {Eardley}, \citenamefont {Lee},
  \citenamefont {Lightman}, \citenamefont {Wagoner},\ and\ \citenamefont
  {Will}}]{Eardley1973a}%
  \BibitemOpen
  \bibfield  {author} {\bibinfo {author} {\bibfnamefont {D.~M.}\ \bibnamefont
  {Eardley}}, \bibinfo {author} {\bibfnamefont {D.~L.}\ \bibnamefont {Lee}},
  \bibinfo {author} {\bibfnamefont {A.~P.}\ \bibnamefont {Lightman}}, \bibinfo
  {author} {\bibfnamefont {R.~V.}\ \bibnamefont {Wagoner}}, \ and\ \bibinfo
  {author} {\bibfnamefont {C.~M.}\ \bibnamefont {Will}},\ }\href {\doibase
  10.1103/PhysRevLett.30.884} {\bibfield  {journal} {\bibinfo  {journal} {Phys.
  Rev. Lett.}\ }\textbf {\bibinfo {volume} {30}},\ \bibinfo {pages} {884}
  (\bibinfo {year} {1973}{\natexlab{a}})}\BibitemShut {NoStop}%
\bibitem [{\citenamefont {Eardley}\ \emph
  {et~al.}(1973{\natexlab{b}})\citenamefont {Eardley}, \citenamefont {Lee},\
  and\ \citenamefont {Lightman}}]{Eardley1973b}%
  \BibitemOpen
  \bibfield  {author} {\bibinfo {author} {\bibfnamefont {D.}~\bibnamefont
  {Eardley}}, \bibinfo {author} {\bibfnamefont {D.}~\bibnamefont {Lee}}, \ and\
  \bibinfo {author} {\bibfnamefont {A.}~\bibnamefont {Lightman}},\ }\href
  {\doibase 10.1103/PhysRevD.8.3308} {\bibfield  {journal} {\bibinfo  {journal}
  {Phys. Rev. D}\ }\textbf {\bibinfo {volume} {8}},\ \bibinfo {pages} {3308}
  (\bibinfo {year} {1973}{\natexlab{b}})}\BibitemShut {NoStop}%
\bibitem [{\citenamefont {Isi}\ \emph {et~al.}(2017)\citenamefont {Isi},
  \citenamefont {Pitkin},\ and\ \citenamefont {Weinstein}}]{Isi2017}%
  \BibitemOpen
  \bibfield  {author} {\bibinfo {author} {\bibfnamefont {M.}~\bibnamefont
  {Isi}}, \bibinfo {author} {\bibfnamefont {M.}~\bibnamefont {Pitkin}}, \ and\
  \bibinfo {author} {\bibfnamefont {A.~J.}\ \bibnamefont {Weinstein}},\ }\href
  {\doibase 10.1103/PhysRevD.96.042001} {\bibfield  {journal} {\bibinfo
  {journal} {Phys. Rev. D}\ }\textbf {\bibinfo {volume} {96}},\ \bibinfo
  {pages} {042001} (\bibinfo {year} {2017})},\ \Eprint
  {http://arxiv.org/abs/1703.07530} {arXiv:1703.07530} \BibitemShut {NoStop}%
\bibitem [{\citenamefont {Callister}\ \emph {et~al.}(2017)\citenamefont
  {Callister}, \citenamefont {Biscoveanu}, \citenamefont {Christensen},
  \citenamefont {Isi}, \citenamefont {Matas}, \citenamefont {Minazzoli},
  \citenamefont {Regimbau}, \citenamefont {Sakellariadou}, \citenamefont
  {Tasson},\ and\ \citenamefont {Thrane}}]{Callister2017}%
  \BibitemOpen
  \bibfield  {author} {\bibinfo {author} {\bibfnamefont {T.}~\bibnamefont
  {Callister}}, \bibinfo {author} {\bibfnamefont {A.~S.}\ \bibnamefont
  {Biscoveanu}}, \bibinfo {author} {\bibfnamefont {N.}~\bibnamefont
  {Christensen}}, \bibinfo {author} {\bibfnamefont {M.}~\bibnamefont {Isi}},
  \bibinfo {author} {\bibfnamefont {A.}~\bibnamefont {Matas}}, \bibinfo
  {author} {\bibfnamefont {O.}~\bibnamefont {Minazzoli}}, \bibinfo {author}
  {\bibfnamefont {T.}~\bibnamefont {Regimbau}}, \bibinfo {author}
  {\bibfnamefont {M.}~\bibnamefont {Sakellariadou}}, \bibinfo {author}
  {\bibfnamefont {J.}~\bibnamefont {Tasson}}, \ and\ \bibinfo {author}
  {\bibfnamefont {E.}~\bibnamefont {Thrane}},\ }\href
  {http://arxiv.org/abs/1704.08373} {\enquote {\bibinfo {title} {{Tests of
  General Relativity with the Stochastic Gravitational-Wave Background}},}\ }
  (\bibinfo {year} {2017}),\ \Eprint {http://arxiv.org/abs/1704.08373}
  {arXiv:1704.08373} \BibitemShut {NoStop}%
\bibitem [{\citenamefont {Thorne}\ \emph {et~al.}(1973)\citenamefont {Thorne},
  \citenamefont {Lee},\ and\ \citenamefont {Lightman}}]{Thorne1973}%
  \BibitemOpen
  \bibfield  {author} {\bibinfo {author} {\bibfnamefont {K.~S.}\ \bibnamefont
  {Thorne}}, \bibinfo {author} {\bibfnamefont {D.~L.}\ \bibnamefont {Lee}}, \
  and\ \bibinfo {author} {\bibfnamefont {A.~P.}\ \bibnamefont {Lightman}},\
  }\href {\doibase 10.1103/PhysRevD.7.3563} {\bibfield  {journal} {\bibinfo
  {journal} {Phys. Rev. D}\ }\textbf {\bibinfo {volume} {7}},\ \bibinfo {pages}
  {3563} (\bibinfo {year} {1973})}\BibitemShut {NoStop}%
\bibitem [{\citenamefont {Brans}\ and\ \citenamefont
  {Dicke}(1961)}]{Brans1961}%
  \BibitemOpen
  \bibfield  {author} {\bibinfo {author} {\bibfnamefont {C.}~\bibnamefont
  {Brans}}\ and\ \bibinfo {author} {\bibfnamefont {R.~H.}\ \bibnamefont
  {Dicke}},\ }\href {\doibase 10.1103/PhysRev.124.925} {\bibfield  {journal}
  {\bibinfo  {journal} {Phys. Rev.}\ }\textbf {\bibinfo {volume} {124}},\
  \bibinfo {pages} {925} (\bibinfo {year} {1961})}\BibitemShut {NoStop}%
\bibitem [{\citenamefont {Andriot}\ and\ \citenamefont
  {G{\'{o}}mez}(2017)}]{Andriot2017}%
  \BibitemOpen
  \bibfield  {author} {\bibinfo {author} {\bibfnamefont {D.}~\bibnamefont
  {Andriot}}\ and\ \bibinfo {author} {\bibfnamefont {G.~L.}\ \bibnamefont
  {G{\'{o}}mez}},\ }\href {\doibase 10.1088/1475-7516/2017/06/048} {\bibfield
  {journal} {\bibinfo  {journal} {J. Cosmol. Astropart. Phys.}\ }\textbf
  {\bibinfo {volume} {2017}},\ \bibinfo {pages} {048} (\bibinfo {year}
  {2017})},\ \Eprint {http://arxiv.org/abs/1704.07392} {arXiv:1704.07392}
  \BibitemShut {NoStop}%
\bibitem [{\citenamefont {Lightman}\ and\ \citenamefont
  {Lee}(1973)}]{Lightman1973}%
  \BibitemOpen
  \bibfield  {author} {\bibinfo {author} {\bibfnamefont {A.~P.}\ \bibnamefont
  {Lightman}}\ and\ \bibinfo {author} {\bibfnamefont {D.~L.}\ \bibnamefont
  {Lee}},\ }\href {\doibase 10.1103/PhysRevD.8.3293} {\bibfield  {journal}
  {\bibinfo  {journal} {Phys. Rev. D}\ }\textbf {\bibinfo {volume} {8}},\
  \bibinfo {pages} {3293} (\bibinfo {year} {1973})}\BibitemShut {NoStop}%
\bibitem [{\citenamefont {Rosen}(1974)}]{Rosen1974}%
  \BibitemOpen
  \bibfield  {author} {\bibinfo {author} {\bibfnamefont {N.}~\bibnamefont
  {Rosen}},\ }\href {\doibase 10.1016/0003-4916(74)90311-X} {\bibfield
  {journal} {\bibinfo  {journal} {Ann. Phys. (N. Y).}\ }\textbf {\bibinfo
  {volume} {84}},\ \bibinfo {pages} {455} (\bibinfo {year} {1974})}\BibitemShut
  {NoStop}%
\bibitem [{\citenamefont {de~Rham}(2014)}]{DeRham2014}%
  \BibitemOpen
  \bibfield  {author} {\bibinfo {author} {\bibfnamefont {C.}~\bibnamefont
  {de~Rham}},\ }\href {\doibase 10.12942/lrr-2014-7} {\bibfield  {journal}
  {\bibinfo  {journal} {Living Rev. Relativ.}\ }\textbf {\bibinfo {volume}
  {17}} (\bibinfo {year} {2014}),\ 10.12942/lrr-2014-7},\ \Eprint
  {http://arxiv.org/abs/1401.4173} {arXiv:1401.4173} \BibitemShut {NoStop}%
\bibitem [{\citenamefont {Mead}(2015)}]{Mead2015}%
  \BibitemOpen
  \bibfield  {author} {\bibinfo {author} {\bibfnamefont {C.}~\bibnamefont
  {Mead}},\ }\href {http://arxiv.org/abs/1503.04866} {\enquote {\bibinfo
  {title} {{Gravitational Waves in G4v}},}\ } (\bibinfo {year} {2015}),\
  \Eprint {http://arxiv.org/abs/1503.04866} {arXiv:1503.04866} \BibitemShut
  {NoStop}%
\bibitem [{\citenamefont {Nishizawa}\ \emph {et~al.}(2009)\citenamefont
  {Nishizawa}, \citenamefont {Taruya}, \citenamefont {Hayama}, \citenamefont
  {Kawamura},\ and\ \citenamefont {Sakagami}}]{Nishizawa2009}%
  \BibitemOpen
  \bibfield  {author} {\bibinfo {author} {\bibfnamefont {A.}~\bibnamefont
  {Nishizawa}}, \bibinfo {author} {\bibfnamefont {A.}~\bibnamefont {Taruya}},
  \bibinfo {author} {\bibfnamefont {K.}~\bibnamefont {Hayama}}, \bibinfo
  {author} {\bibfnamefont {S.}~\bibnamefont {Kawamura}}, \ and\ \bibinfo
  {author} {\bibfnamefont {M.-a.}\ \bibnamefont {Sakagami}},\ }\href {\doibase
  10.1103/PhysRevD.79.082002} {\bibfield  {journal} {\bibinfo  {journal} {Phys.
  Rev. D}\ }\textbf {\bibinfo {volume} {79}},\ \bibinfo {pages} {082002}
  (\bibinfo {year} {2009})},\ \Eprint {http://arxiv.org/abs/0903.0528}
  {arXiv:0903.0528} \BibitemShut {NoStop}%
\bibitem [{\citenamefont {B{\l}aut}(2012)}]{Blaut2012}%
  \BibitemOpen
  \bibfield  {author} {\bibinfo {author} {\bibfnamefont {A.}~\bibnamefont
  {B{\l}aut}},\ }\href {\doibase 10.1103/PhysRevD.85.043005} {\bibfield
  {journal} {\bibinfo  {journal} {Phys. Rev. D}\ }\textbf {\bibinfo {volume}
  {85}},\ \bibinfo {pages} {043005} (\bibinfo {year} {2012})}\BibitemShut
  {NoStop}%
\bibitem [{\citenamefont {Isi}\ \emph {et~al.}(2015)\citenamefont {Isi},
  \citenamefont {Weinstein}, \citenamefont {Mead},\ and\ \citenamefont
  {Pitkin}}]{Isi2015}%
  \BibitemOpen
  \bibfield  {author} {\bibinfo {author} {\bibfnamefont {M.}~\bibnamefont
  {Isi}}, \bibinfo {author} {\bibfnamefont {A.~J.}\ \bibnamefont {Weinstein}},
  \bibinfo {author} {\bibfnamefont {C.}~\bibnamefont {Mead}}, \ and\ \bibinfo
  {author} {\bibfnamefont {M.}~\bibnamefont {Pitkin}},\ }\href {\doibase
  10.1103/PhysRevD.91.082002} {\bibfield  {journal} {\bibinfo  {journal} {Phys.
  Rev. D}\ }\textbf {\bibinfo {volume} {91}},\ \bibinfo {pages} {082002}
  (\bibinfo {year} {2015})}\BibitemShut {NoStop}%
\bibitem [{\citenamefont {Poisson}\ and\ \citenamefont
  {Will}(2014)}]{Poisson2014}%
  \BibitemOpen
  \bibfield  {author} {\bibinfo {author} {\bibfnamefont {E.}~\bibnamefont
  {Poisson}}\ and\ \bibinfo {author} {\bibfnamefont {C.~M.}\ \bibnamefont
  {Will}},\ }\href@noop {} {\emph {\bibinfo {title} {{Gravity: Newtonian,
  Post-Newtonian, Relativistic}}}}\ (\bibinfo  {publisher} {Cambridge
  University Press},\ \bibinfo {address} {Cambridge},\ \bibinfo {year}
  {2014})\BibitemShut {NoStop}%
\bibitem [{\citenamefont {Lee}\ \emph {et~al.}(2008)\citenamefont {Lee},
  \citenamefont {Jenet},\ and\ \citenamefont {Price}}]{Lee2008}%
  \BibitemOpen
  \bibfield  {author} {\bibinfo {author} {\bibfnamefont {K.~J.}\ \bibnamefont
  {Lee}}, \bibinfo {author} {\bibfnamefont {F.~A.}\ \bibnamefont {Jenet}}, \
  and\ \bibinfo {author} {\bibfnamefont {R.~H.}\ \bibnamefont {Price}},\ }\href
  {\doibase 10.1086/591080} {\bibfield  {journal} {\bibinfo  {journal}
  {Astrophys. J.}\ }\textbf {\bibinfo {volume} {685}},\ \bibinfo {pages} {1304}
  (\bibinfo {year} {2008})}\BibitemShut {NoStop}%
\bibitem [{\citenamefont {Chamberlin}\ and\ \citenamefont
  {Siemens}(2012)}]{Chamberlin2012}%
  \BibitemOpen
  \bibfield  {author} {\bibinfo {author} {\bibfnamefont {S.~J.}\ \bibnamefont
  {Chamberlin}}\ and\ \bibinfo {author} {\bibfnamefont {X.}~\bibnamefont
  {Siemens}},\ }\href {\doibase 10.1103/PhysRevD.85.082001} {\bibfield
  {journal} {\bibinfo  {journal} {Phys. Rev. D}\ }\textbf {\bibinfo {volume}
  {85}},\ \bibinfo {pages} {082001} (\bibinfo {year} {2012})},\ \Eprint
  {http://arxiv.org/abs/1111.5661} {arXiv:1111.5661} \BibitemShut {NoStop}%
\bibitem [{\citenamefont {Cornish}\ and\ \citenamefont
  {Littenberg}(2014)}]{Cornish2014}%
  \BibitemOpen
  \bibfield  {author} {\bibinfo {author} {\bibfnamefont {N.~J.}\ \bibnamefont
  {Cornish}}\ and\ \bibinfo {author} {\bibfnamefont {T.~B.}\ \bibnamefont
  {Littenberg}},\ }\href {\doibase 10.1088/0264-9381/32/13/135012} {\bibfield
  {journal} {\bibinfo  {journal} {Class. Quantum Gravity}\ }\textbf {\bibinfo
  {volume} {32}},\ \bibinfo {pages} {135012} (\bibinfo {year} {2014})},\
  \Eprint {http://arxiv.org/abs/1410.3835} {arXiv:1410.3835} \BibitemShut
  {NoStop}%
\bibitem [{\citenamefont {Littenberg}\ \emph {et~al.}(2016)\citenamefont
  {Littenberg}, \citenamefont {Kanner}, \citenamefont {Cornish},\ and\
  \citenamefont {Millhouse}}]{Littenberg2016}%
  \BibitemOpen
  \bibfield  {author} {\bibinfo {author} {\bibfnamefont {T.~B.}\ \bibnamefont
  {Littenberg}}, \bibinfo {author} {\bibfnamefont {J.~B.}\ \bibnamefont
  {Kanner}}, \bibinfo {author} {\bibfnamefont {N.~J.}\ \bibnamefont {Cornish}},
  \ and\ \bibinfo {author} {\bibfnamefont {M.}~\bibnamefont {Millhouse}},\
  }\href {\doibase 10.1103/PhysRevD.94.044050} {\bibfield  {journal} {\bibinfo
  {journal} {Phys. Rev. D}\ }\textbf {\bibinfo {volume} {94}},\ \bibinfo
  {pages} {044050} (\bibinfo {year} {2016})},\ \Eprint
  {http://arxiv.org/abs/1511.08752} {arXiv:1511.08752} \BibitemShut {NoStop}%
\bibitem [{\citenamefont {Veitch}\ \emph {et~al.}(2015)\citenamefont {Veitch},
  \citenamefont {Raymond}, \citenamefont {Farr}, \citenamefont {Farr},
  \citenamefont {Graff}, \citenamefont {Vitale}, \citenamefont {Aylott},
  \citenamefont {Blackburn}, \citenamefont {Christensen}, \citenamefont
  {Coughlin}, \citenamefont {{Del Pozzo}}, \citenamefont {Feroz}, \citenamefont
  {Gair}, \citenamefont {Haster}, \citenamefont {Kalogera}, \citenamefont
  {Littenberg}, \citenamefont {Mandel}, \citenamefont {O'Shaughnessy},
  \citenamefont {Pitkin}, \citenamefont {Rodriguez}, \citenamefont
  {R{\"{o}}ver}, \citenamefont {Sidery}, \citenamefont {Smith}, \citenamefont
  {{Van Der Sluys}}, \citenamefont {Vecchio}, \citenamefont {Vousden},\ and\
  \citenamefont {Wade}}]{Veitch2015}%
  \BibitemOpen
  \bibfield  {author} {\bibinfo {author} {\bibfnamefont {J.}~\bibnamefont
  {Veitch}}, \bibinfo {author} {\bibfnamefont {V.}~\bibnamefont {Raymond}},
  \bibinfo {author} {\bibfnamefont {B.}~\bibnamefont {Farr}}, \bibinfo {author}
  {\bibfnamefont {W.}~\bibnamefont {Farr}}, \bibinfo {author} {\bibfnamefont
  {P.}~\bibnamefont {Graff}}, \bibinfo {author} {\bibfnamefont
  {S.}~\bibnamefont {Vitale}}, \bibinfo {author} {\bibfnamefont
  {B.}~\bibnamefont {Aylott}}, \bibinfo {author} {\bibfnamefont
  {K.}~\bibnamefont {Blackburn}}, \bibinfo {author} {\bibfnamefont
  {N.}~\bibnamefont {Christensen}}, \bibinfo {author} {\bibfnamefont
  {M.}~\bibnamefont {Coughlin}}, \bibinfo {author} {\bibfnamefont
  {W.}~\bibnamefont {{Del Pozzo}}}, \bibinfo {author} {\bibfnamefont
  {F.}~\bibnamefont {Feroz}}, \bibinfo {author} {\bibfnamefont
  {J.}~\bibnamefont {Gair}}, \bibinfo {author} {\bibfnamefont {C.-J.}\
  \bibnamefont {Haster}}, \bibinfo {author} {\bibfnamefont {V.}~\bibnamefont
  {Kalogera}}, \bibinfo {author} {\bibfnamefont {T.}~\bibnamefont
  {Littenberg}}, \bibinfo {author} {\bibfnamefont {I.}~\bibnamefont {Mandel}},
  \bibinfo {author} {\bibfnamefont {R.}~\bibnamefont {O'Shaughnessy}}, \bibinfo
  {author} {\bibfnamefont {M.}~\bibnamefont {Pitkin}}, \bibinfo {author}
  {\bibfnamefont {C.}~\bibnamefont {Rodriguez}}, \bibinfo {author}
  {\bibfnamefont {C.}~\bibnamefont {R{\"{o}}ver}}, \bibinfo {author}
  {\bibfnamefont {T.}~\bibnamefont {Sidery}}, \bibinfo {author} {\bibfnamefont
  {R.}~\bibnamefont {Smith}}, \bibinfo {author} {\bibfnamefont
  {M.}~\bibnamefont {{Van Der Sluys}}}, \bibinfo {author} {\bibfnamefont
  {A.}~\bibnamefont {Vecchio}}, \bibinfo {author} {\bibfnamefont
  {W.}~\bibnamefont {Vousden}}, \ and\ \bibinfo {author} {\bibfnamefont
  {L.}~\bibnamefont {Wade}},\ }\href {\doibase 10.1103/PhysRevD.91.042003}
  {\bibfield  {journal} {\bibinfo  {journal} {Phys. Rev. D}\ }\textbf {\bibinfo
  {volume} {91}},\ \bibinfo {pages} {042003} (\bibinfo {year} {2015})},\
  \Eprint {http://arxiv.org/abs/1409.7215} {arXiv:1409.7215} \BibitemShut
  {NoStop}%
\end{thebibliography}%

\end{document}